# Designs for programmable quantum resistance standards based on epitaxial graphene *p-n* junctions

Jiuning Hu[1,2,*], Albert F. Rigosi[1*], Mattias Kruskopf[1,2], Yanfei Yang[1,2], Bi-Yi Wu[1,3], Jifa Tian[1,4], Alireza R. Panna[1], Hsin-Yen Lee[1,5], Shamith U. Payagala[1], George R. Jones[1], Marlin E. Kraft[1], Dean G. Jarrett[1], Kenji Watanabe[6], Takashi Taniguchi[6], Randolph E. Elmquist[1], and David B. Newell[1]

[1]Physical Measurement Laboratory, National Institute of Standards and Technology (NIST), Gaithersburg, MD 20899, USA
[2]Joint Quantum Institute, University of Maryland, College Park, MD 20742, USA
[3]Graduate Institute of Applied Physics, National Taiwan University, Taipei 10617, Taiwan
[4]Department of Physics and Astronomy, and Birck Nanotechnology Center, Purdue University, West Lafayette, Indiana 47907, USA
[5]Theiss Research, La Jolla, CA 92037, USA
[6]National Institute for Materials Science, 1-1 Namiki, Tsukuba 305-0044, Japan
*hujiuning@gmail.com

## Abstract

We report the fabrication and measurement of top gated epitaxial graphene *p-n* junctions where exfoliated hexagonal boron nitride (*h*-BN) is used as the gate dielectric. The four-terminal longitudinal resistance across a single junction is well quantized at the von Klitzing constant $R_K$ with a relative uncertainty of $10^{-7}$. After the exploration of numerous parameter spaces, we summarize the conditions upon which these devices could function as potential resistance standards. Furthermore, we offer designs of programmable electrical resistance standards over six orders of magnitude by using external gating.

## Introduction

The advantageous electrical and optical properties of graphene have been well-studied.[1-4] When compared to semiconductor heterostructures, epitaxial graphene (EG) on silicon carbide (SiC) has been identified as an ideal platform for resistance standards due to the observation of the quantum Hall effect (QHE) with resistance plateaus that span over a wide range of magnetic flux densities, large breakdown currents, and operation at relatively high temperatures.[5-13] These resistance standards exclusively operate at the filling factor $\nu = 2$, corresponding to the resistance value: $\frac{1}{2}\frac{h}{e^2} = \frac{1}{2}R_K$ [see additional information], where $h$ is Planck's constant and $e$ is the elementary charge. Moreover, the realization of other values based on fundamental constants is a crucial milestone in resistance metrology that is still being explored with EG. One approach is to connect multiple Hall bars in complicated parallel and series networks to create resistance values of $qR_K$ where $q$ is a positive rational number.[14-16] The other approach is to utilize the unique properties of graphene to build p-n junctions working in the quantum Hall regime that allow convenient resistance scaling.[17]

Due to its linear dispersion relation, where the characteristic Dirac point represents charge neutrality, it is easy to electronically dope graphene into bipolar carrier concentration regions, denoted as $p$ (holes) or $n$ (electrons), with external gates. Furthermore, the physics of graphene $p$-$n$ junctions ($pn$Js) enables one to fabricate devices to access quantized resistance values that are multiples or fractions of $\frac{1}{2}R_K$.[17] When all of the regions of a device are in the quantum Hall regime but have different carrier concentrations and polarities, the measured longitudinal resistivities across one or several sets of $pn$Js depend on how the Landauer-Büttiker edge states equilibrate at the junction.[18-20] There have been several reports about this behavior while using tunable gates to adjust the $pn$J.[18, 21-24]

Graphene $pn$J can be utilized to circumvent frequent technical difficulties resulting from the general use of metallic contacts and multiple device interconnections. In this work, we demonstrate highly accurate resistance quantization at $R_K$ in an EG $pn$J device, measured with a direct current comparator (DCC) resistance bridge. The characterization measurements are summarized here, starting with a description of the device engineering. Comprehensive exploration of experimental parameter spaces (B field, gate voltages, temperature) was performed, enabling us to fully understand this EG $pn$J, and further motivating us to discuss the conceptual realization of constructing a programmable quantized Hall resistance (PQHR) system.

## Results

The EG Hall bar is shown in Figure 1 (a) indicated by the red dashed line with a width of 50 μm and the blue dashed line showing the perimeter of the $h$-BN flake, measured to have a thickness of 45 nm ($d_{BN}$). The atomic force microscope (AFM) image in Figure 1 (b) is a magnification of one of the top gate junctions, represented by the small black square in (a), with dashed gray lines used to help identify that region. The width of the gap between the two gates is approximately 150 nm, which is small enough to create a single $pn$J. A cross-section illustration of the device is shown in Figure 1 (c) depicting the use of $h$-BN as a dielectric layer. The top gate quality was assessed by measuring the leakage current between top gate and EG. When a DC voltage was applied between graphene and the top gates G1 and G2 through the range of -15 V to 15 V, a leakage current of less than 1 nA was observed. For voltages that exceeded 15 V in either polarity, the leakage current rapidly increased beyond 1 nA, thus defining an upper and lower bound to the gate voltage magnitude. Due to excessive leakage currents, gate G3 in Fig. 1 (a) was not used.

One would ideally like to know the range of carrier densities attainable with these gates, and so we first focus on determining the carrier density parameter space in the unipolar case. The well-documented electrical and optical properties of the interfacial buffer layer as well as its interactions with EG allow one to expect inherent *n*-type doping.[25-27] We employ a basic capacitance model to gain an insight into the expected doping. Let $n_G$ and $E_F$ be the electron density and Fermi level of the EG layer, respectively. The relationship between those two parameters and the gate voltage $V_G$ is:[28-29]

$$\frac{C_{ox}}{e}(V_G - V_D) = \frac{C_{y2}}{C_{s2}}\left[n_G + (C_{s1} + C_{s2})\frac{E_F}{e^2}\right] \qquad (1)$$

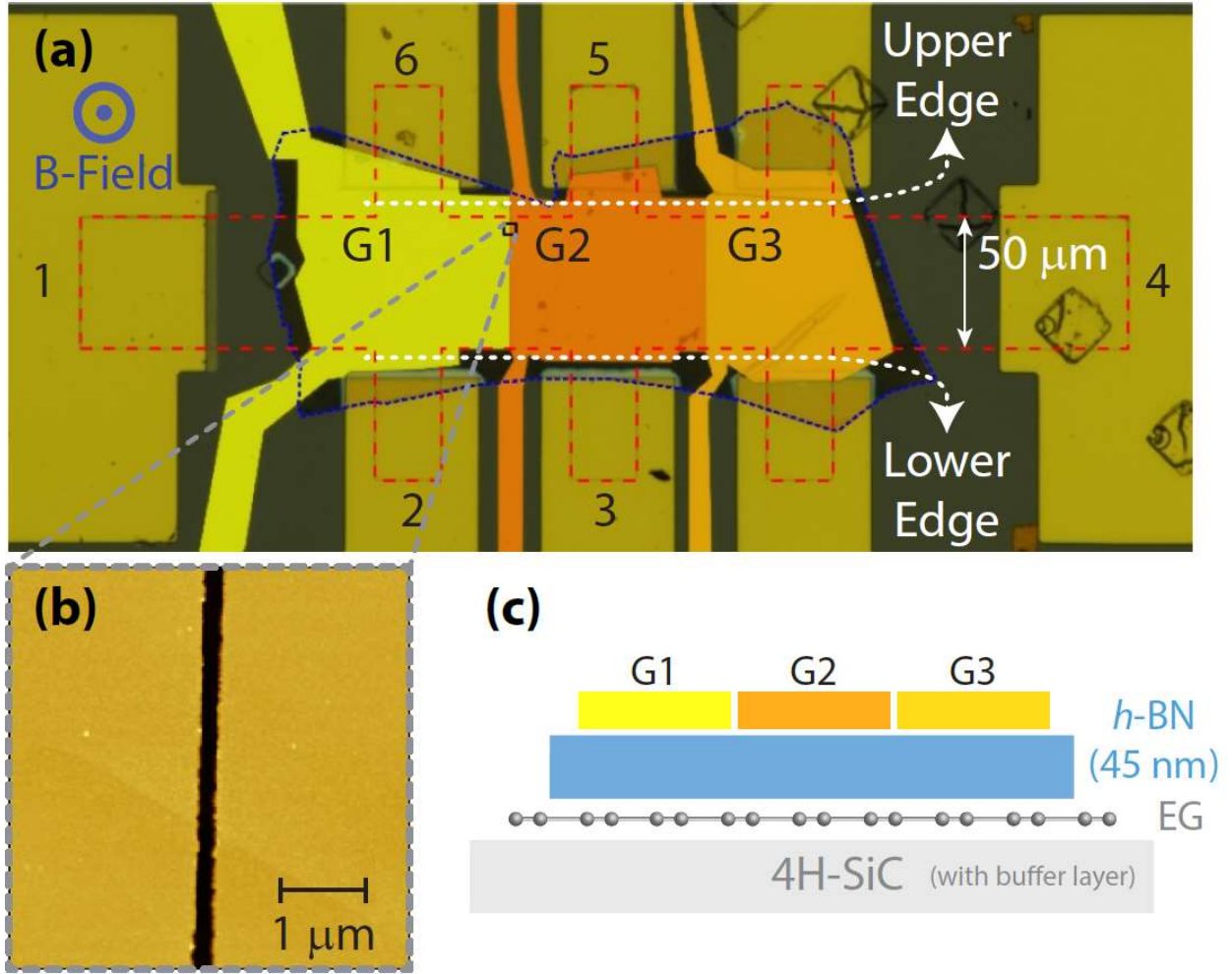

**Figure 1. EG device description.** (**a**) The optical image of the device shows the graphene Hall bar outlined by a red dashed line and the encapsulating *h*-BN layer enclosed by a blue dotted line. The width of the Hall bar channel is marked with a white arrow. Each used device contact is given a numerical designation and the top gates are labelled as G1, G2 and G3 and represented as digitally artificial colors of yellow, orange, and bronze, respectively. The upper (UE) and lower edges (LE) are marked with dashed white lines. The direction of the positive magnetic field for the measurements is out of the page. (**b**) The atomic force microscope image of the *pn*J shows the magnified region enclosed by the black square in (**a**). (**c**) An illustration device's cross section is provided.

In equation (1), $V_D$ is the voltage corresponding to the Dirac point, $C_{ox} = \frac{\epsilon_{BN}\epsilon_0}{d_{BN}}$ is the gate geometric capacitance (per unit area), with $\epsilon_{BN}$ and $d_{BN}$ as the dielectric constant and thickness of *h*-BN, respectively. The other constants are determined by the quantum capacitance of the charge transfer layers including EG, the buffer layer beneath, and the residual chemical doping between graphene and *h*-BN. Additionally, $C_s$ is the additive capacitance of the quantum capacitance and the single layer gap of 0.3 nm between graphene and the adjacent layer.

The capacitance relations are listed here, with $d = 0.3$ nm as the distance between the EG and its adjacent neighboring layers, where the buffer is labelled by subscript $i = 1$ and residual chemical doping is labelled by subscript $i = 2$ such that: (1) quantum capacitance $C_{\gamma i} = \gamma_i e^2$ (2) geometrical capacitance $C_{ci} = \frac{\epsilon_i}{d}$ and total capacitance (3) $C_{si} = \left( \frac{1}{C_{ci}} + \frac{1}{C_{\gamma i}} \right)^{-1}$. The used dielectric constants are $\epsilon_{BN} = 3.9\epsilon_0$, $\epsilon_1 = 9.7\epsilon_0$, and $\epsilon_2 = 3\epsilon_0$, where $\epsilon_0$ is the vacuum permittivity.[28-29] In the case of zero magnetic field, the relation between $n_G$ and $E_F$ is $E_F = \hbar v_F \sqrt{\pi |n_G|} sign(n_G)$, while in nonzero magnetic field B, the total density can be found with $n_G = \int_0^{E_F} D(E) dE$, with the density of states:[30]

$$D(E) = \frac{g_s g_v eB}{h} \sum_N \frac{1}{\sqrt{2\pi s^2}} exp \left[ \frac{-(E - E_N)^2}{2s^2} \right] \qquad (2)$$

where $E_N = v_F \sqrt{2|N|} sign(N)$ is the Landau level $N$, $s = 12$ meV, $g_s g_v = 2$ are the spin and valley degeneracies.

Equation (2) gives the total density of states (DOS), which, when combined with equation (1), allows us to determine how $n_G$ varies as a function of unipolar gate voltage. This calculation also depends on the capacitance parameters $\gamma_1$ and $\gamma_2$, representing the two interfaces adjacent to the graphene sheet. In the case where $\gamma_1$ and $\gamma_2$ are zero, one assumes a freestanding EG layer with no charge transfer between interfaces. As the two parameters are modulated, one can fit this model to experimental data. Figure 2 (a) shows measurements for $R_{xx}$ and $R_{xy}$, collected at $B = 0.2$ T. Those two quantities are then converted into $n_G$ and $\mu$, as seen in Figure 2 (b) as orange data points and a green curve, respectively. The following formulas were used to calculate $n_G$ and the mobility: $n_G = \frac{1}{e\left(\frac{dR_{xy}}{dB}\right)}$ and $\mu = \frac{1}{en_G R_{xx}\frac{W}{L}}$. The capacitance model is plotted here as well, showing the case where EG is freestanding and the case where the model fits the data for $n_G$, which corresponds to $\gamma_1 = 1.2 \times 10^{14} eV^{-1}cm^{-2}$ and $\gamma_2 = 1.5 \times 10^4 eV^{-1}cm^{-2}$. Note that there are four different ways to calculate $n_G$ and mobility using the data in Figure 2 (a). The results presented by the dashed lines in Figure 2 (b) are the mean of the calculations.

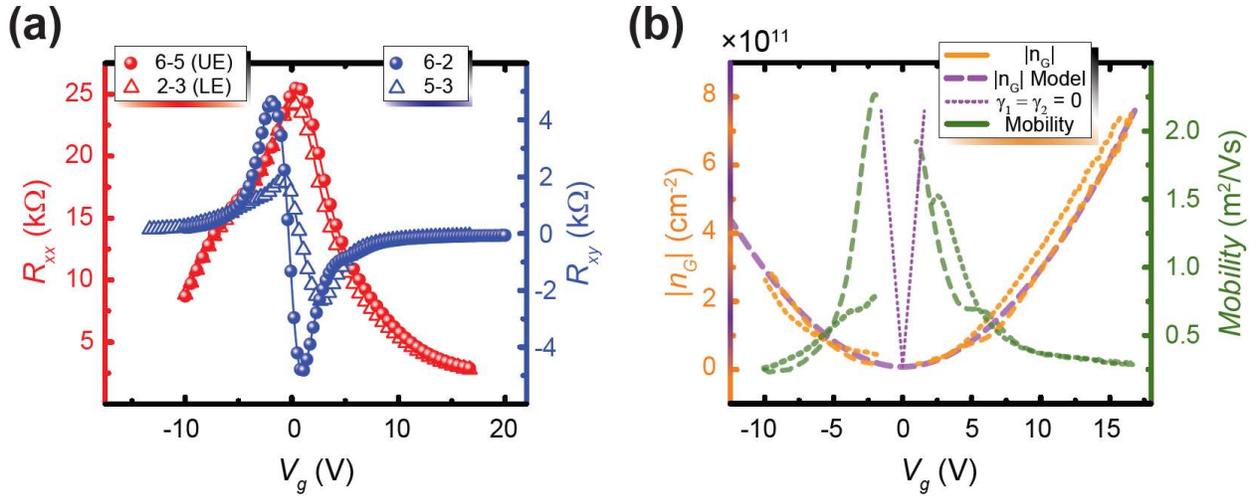

**Figure 2. EG device characterization.** (**a**) At $B$ = 0.2 T and $T$ = 1.7 K, the device's longitudinal resistances ($R_{xx}$) in red and Hall resistances ($R_{xy}$) in blue are measured with respect to unipolar gate voltage. $R_{xx}$ is measured for the upper and lower edge (UE and LE), shown as points and triangles, respectively, whereas the point and triangle symbols for $R_{xy}$ represent just two different regions of the device. All legend numbers correspond to contact numbers from Fig. 1. (**b**) The carrier densities are measured (orange curves, dashed and dotted for the two different regions), the mobilities are calculated (green curve, with dashed and dotted appearances corresponding to the same two regions), and the capacitance models for both freestanding EG and EG undergoing charge transfer (due to adjacent neighboring layers) are plotted (pink dotted and dashed purple curves, respectively). The carrier densities and capacitance models are plotted on the left vertical axis, whereas the mobilities are plotted on the right vertical axis.

With the unipolar bounds of $n_G$ known as about 6 × 10¹¹ cm⁻² at a gate voltage of 15 V, $V_G$ and $B$ were used as independent variables to determine the boundary beyond which full quantization occurs. These measurements, accompanied by Hall resistance measurements, are shown in

Figure 3. In Fig. 3 (a), the dashed lines are calculated from Equation (1) and (2) using the fitted results of to $\gamma_1$ and $\gamma_2$. From these measurements, one observes only two definitive regions of zero longitudinal resistance, which indicates that only the $\nu = \pm 2$ plateau is accurately quantized.

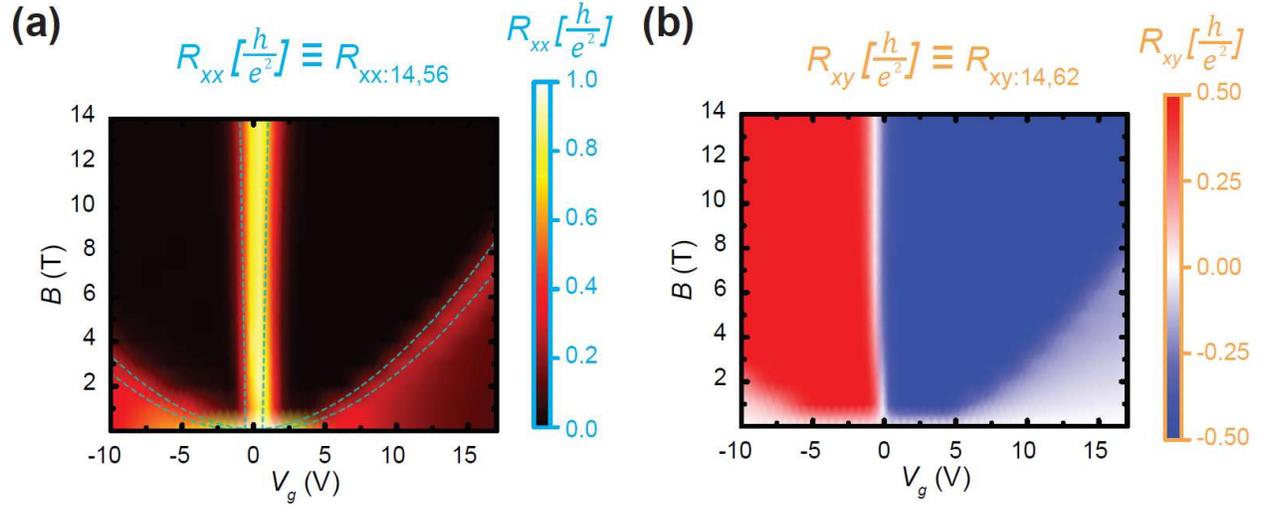

**Figure 3. Conditions for full quantization.** (**a**) At $T = 1.7$ K, $R_{xx}$ for the upper edge is measured as a function of unipolar gate voltage and magnetic field. Contact numbers correspond to those in Fig. 1. A model for determining the boundaries for quantization are marked in dotted cyan. $R_{xx}$ clearly drops to zero past this boundary into the black-colored region. (**b**) $R_{xy}$ measured in the G1 region is provided to verify that only $\nu = \pm 2$ is well-quantized, with the same parabolic boundary appearing in the data.

After determining the magnetic field conditions for full quantization, we explore $R_{xx}$ as a function of the two gate voltages in Figure 4 at $B = 14$ T. The upper edge in Fig. 4 (a) has zero resistance for three out of four quadrants in this parameter space. For the final quadrant, corresponding to G2 and G1 as an $n$-type and $p$-type region, respectively, $R_{xx}$ takes on the quantized value of $R_K$ [see additional information]. Full quantization is further verified by the accompanying Hall resistance measurements for G2 and G1 seen in Fig. 4 (b) and (d). The lower edge in Fig. 4 (c) exhibits the same behavior when the gate polarities are reversed. Furthermore, the data is symmetric about the diagonal cut which intersects the Dirac points of G1 and G2.

More information about the two Dirac point measurements can be seen in the Supplementary information.

 One advantage of these observations of the longitudinal resistance taking on different values is that such values could potentially be used for electrically programmable quantum resistance standards. These observations motivated the continued exploration of how the longitudinal resistance of the *pn*Js can be accurately quantized at $R_K$ under appropriate gate conditions. In order to characterize the device for metrological purposes, critical currents of QHE breakdown must be known. Higher critical currents are generally beneficial for resistance metrology since they can both provide a better signal-to-noise ratio and an increased compatibility with commercial metrology equipment such as the DCC, voltage calibrators, and precision digital multimeters.[31-33]

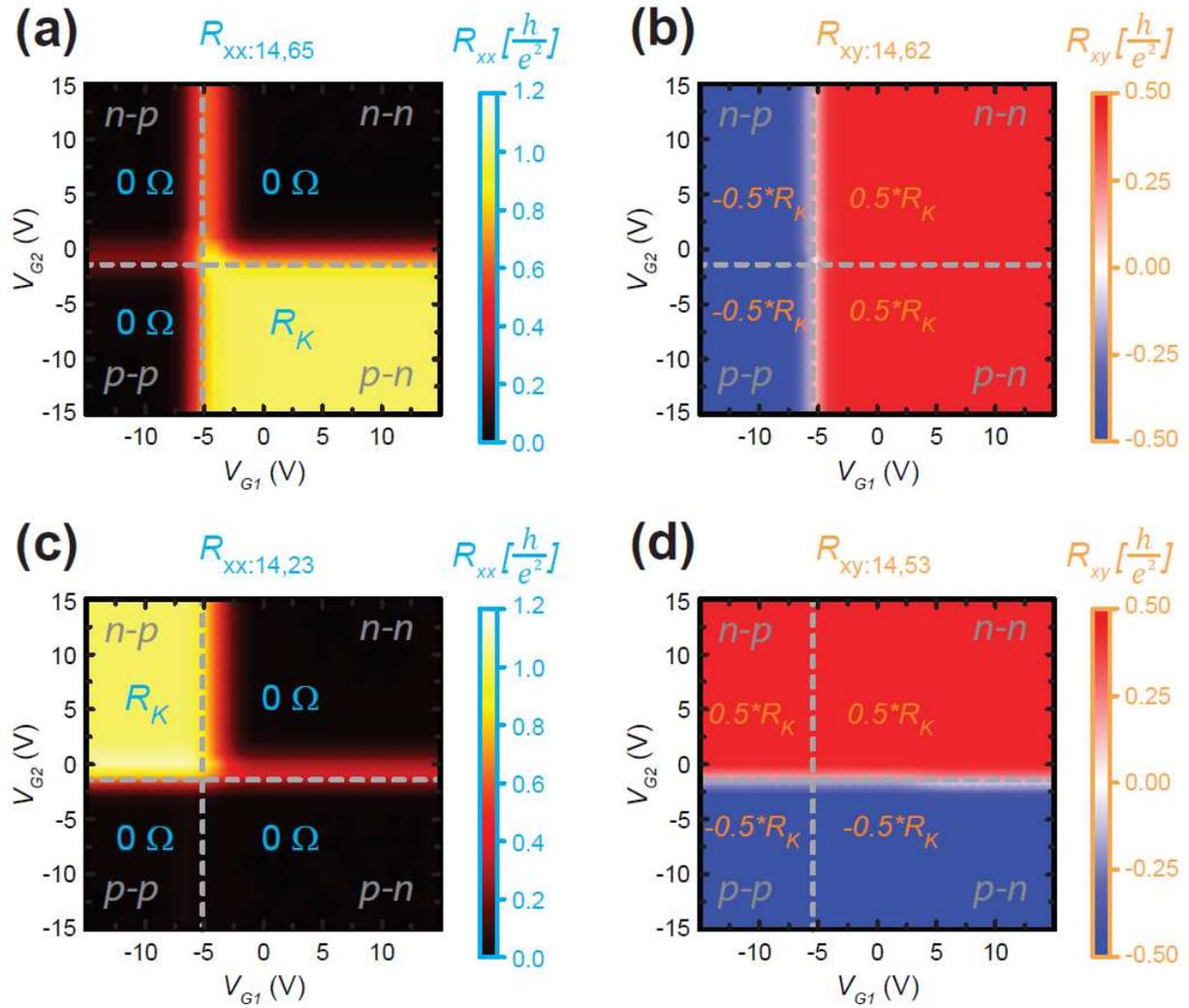

**Figure 4. $R_{xx}$ and $R_{xy}$ as a function of two gate voltages.** (**a**) At $B = 14$ T (as shown in Fig. 1 (a)) and $T = 1.7$ K, $R_{xx}$ for upper edge is measured as a function of two gate voltages. In both unipolar and one bipolar region, $R_{xx}$ is measured across a low dissipation region, and thus takes on the expected value of zero. When the graphene under G2 and G1 is $n$-type and $p$-type, respectively, $R_{xx}$ takes on the quantized value of $R_K$. (**b**) The corresponding Hall resistances for the same parameter space are shown to verify that the device is quantized. (**c**) The lower edge is measured as a function of two gate voltages and expectedly exhibits the same behavior when the polarity of the top gates is reversed. (**d**) With the Hall resistances verifying the quantization in the lower edge case, it becomes clear that all four maps are symmetric about a diagonal line which intersects the Dirac points of both G1 and G2.

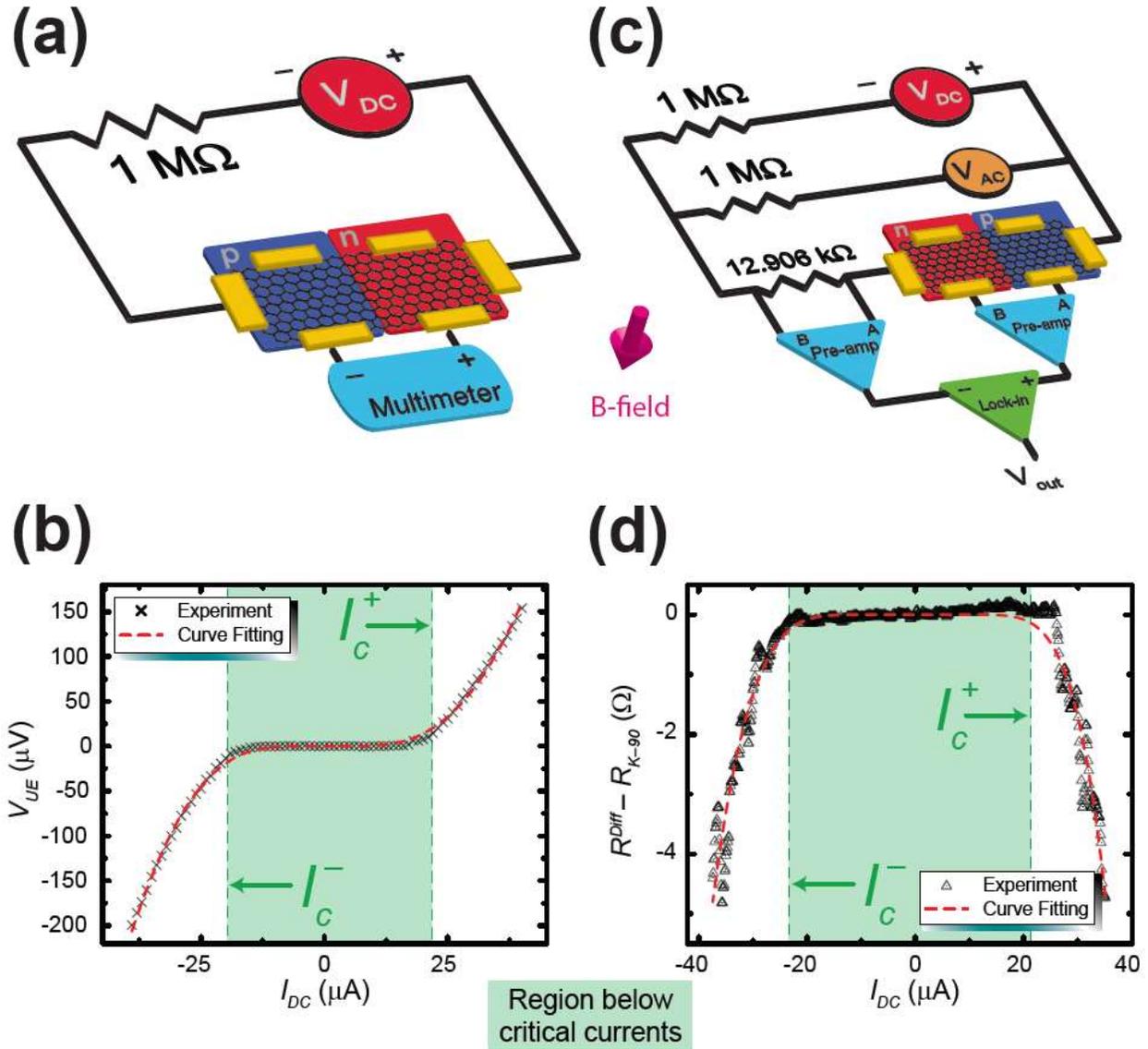

**Figure 5. Measurement techniques for finding the critical currents at different quantized values of $R_{xx}$.** (**a**) A circuit diagram is illustrated for the experimental setup which allows one to determine the critical current bounds of the *pn*J while $R_{xx}$ is characterized by a region of low dissipation (i.e. a region yielding 0 Ω). (**b**) An example measurement for the circuit diagram in (**a**) shows that for a quantized device, zero voltage across the multimeter is expected for small applied currents, and once quantization breaks down, the device contributes a nonzero voltage drop to the measurement. The data points and data fitting curve are shown as black 'X' marks and a dashed red curve, respectively. (**c**) For the case when $R_{xx}$ is quantized at $R_K$, a more intricate experimental setup is required involving the use of a standard resistor in series with the EG *pn*J device to remove the large DC component for the differential measurement. (**d**) The circuit in (**c**) is a direct determination of the resistance, as defined by measuring differential voltage with respect to differential current. $R_K$ is automatically subtracted from the differential resistance to give a region where the subtraction yields zero. The data points and data fitting curve are shown as black triangles and a dashed red curve, respectively. For (**b**) and (**d**), the shaded green area marks a boundary, within which the applied current is still below the critical current.

To measure the critical currents of the entire device as a function of two gate voltages, different circuits required assembly depending on which of the two quantized values of $R_{xx}$ was being measured. Here, the critical current is defined as the point when is less than the relative uncertainty, indicating that the device is no longer exhibiting the QHE with full quantization. This condition will be further clarified later in the text. Figure 5 (a) contains an illustration of the circuit used to measure the critical currents for the device while it is quantized. An example measurement representing how three out of four regions in the dual gate parameter space behave is shown in Fig. 5 (b). In this example case, using $V_{G1} = V_{G2} = 7\ V$, the negative and positive critical current limits are found to be about -21 μA and 24 μA. The measured result can be well-described by the formula of variable range hopping (VRH) transport: $V_{UE} \propto I_{DC} exp\left[-\sqrt{I_0/|I_{DC}|}\right].$[34]

For the case of $R_{xx} = R_K$, a different circuit is required to measure the critical currents, shown in Fig. 5 (c). The circuit essentially allows one to measure the differential resistance at different DC current biases. The measured differential resistance shown in Fig. 5 (d) is approximately zero for the region between -23 μA and 21 μA, and can be fit with the formula: $R_{diff} - R_K \propto \left[1 + \frac{3}{2}\sqrt{I_0/|I_{DC}|}\right] I_{DC} exp\left[-\sqrt{I_0/|I_{DC}|}\right]$. As with Fig. 5 (b), these regions below the critical currents are shaded in green.

The critical currents were then determined as a function of two gate voltages in Figure 6. Visually, one can associate the intensity of a point in these dual gate parameter spaces with the endpoints of the green shaded area in Fig. 5 (b) or (d). For instance, the black region in Fig. 6 (a) corresponds to a value of 0 μA, meaning that the device is not accurately quantized. For the UE, if $V_{G1} = V_{G2} = 10V$ (see Fig. 6 (a) and (b)), the $I_c^-$ and $I_c^+$ limits are about -60 μA and 40 μA, respectively.

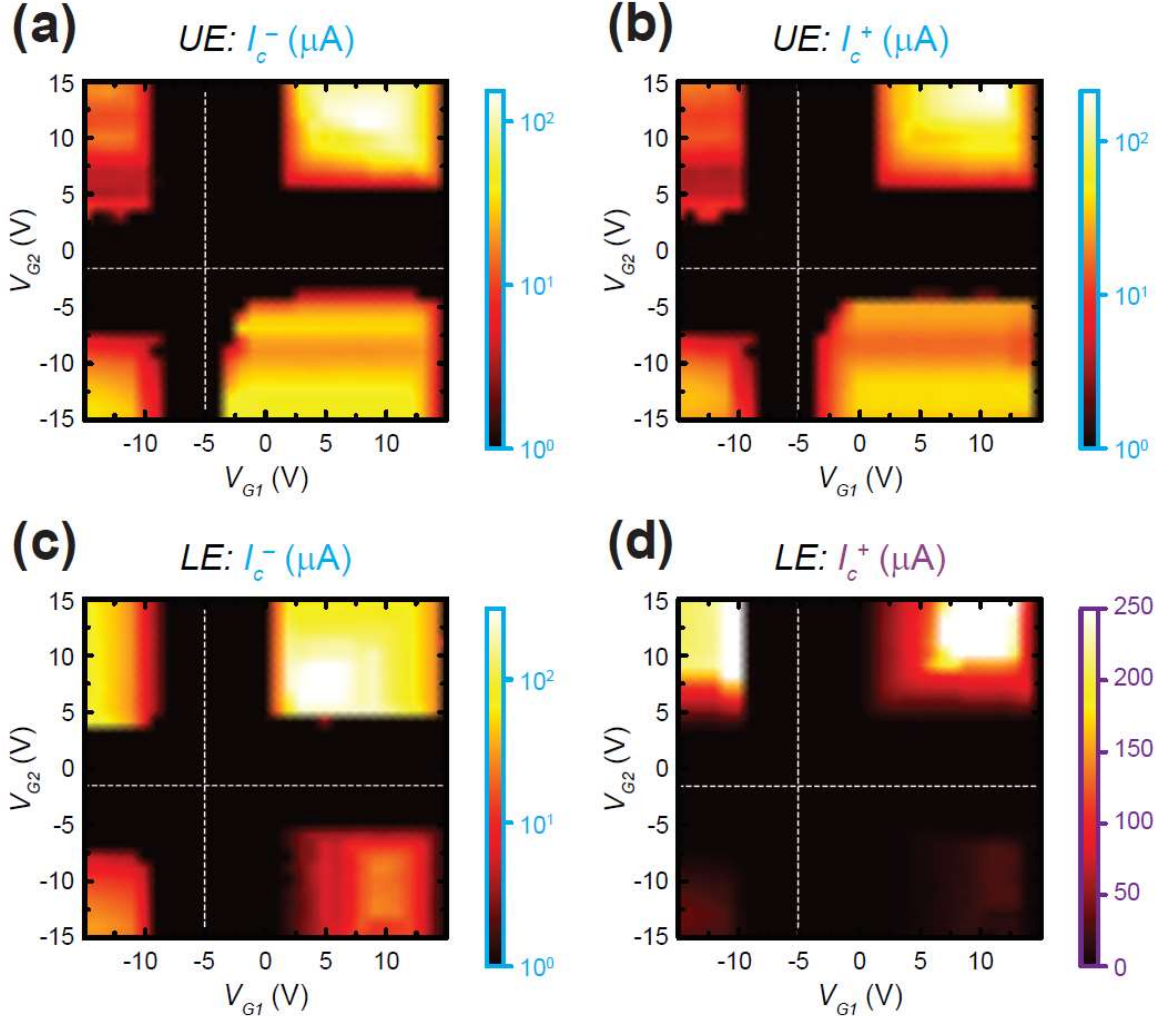

**Figure 6. Critical currents at different gate voltages.** (**a**) At $B = 14$ T and $T = 1.7$ K, the negative limit of critical current $I_c^-$ is determined for the upper edge (UE). The scale is logarithmic in current and represents the data for three quadrants measured with the circuit in Fig. 5 (a), whereas the fourth quadrant (upper left corner) shows data measured with the circuit in Fig. 5 (c). The dotted white lines are a guide to visualize the Dirac points of regions G1 and G2. (**b**) The same type of data as (a) are shown, only representing the positive limit of critical current $I_c^+$. (**c**) Data are shown with the same conditions as (**a**), but with the measurement focusing on the LE of the device. (**d**) $I_c^+$ is determined for the lower edge (LE) in the same fashion as (**b**), with the main difference being the replacement of the logarithmic scale with a linear one to demonstrate the order-of-magnitude difference in critical current limits for certain unipolar or bipolar configurations.

Figure 6 gives a full understanding of the critical current limits $I_c^-$ and $I_c^+$ for the upper edge ((a) and (b)) and the lower edge ((c) and (d)) of the device. It should be noted that for the upper left corner of the maps in (a) and lower right in (b), the measurements were taken using the circuit shown in Fig. 5 (c), and the other three quadrants, with respect to the Dirac points, were measured using the circuit in Fig. 5 (a).

With the critical currents' behavior generally understood for the two gate voltages, one final parameter was tested. For $V_{G1} = 10V$ and $V_{G2} = -10V$, we monitored the lower edge for deviations from the quantized value of $R_K$ and $I_c^-$ and $I_c^+$ were determined as a function of temperature. The overall dependence of the deviations from $R_K$ as a function of current and temperature is presented in Figure 7 (a), and the dependence of the extracted critical currents $I_c^-$ and $I_c^+$ on temperature, for both polarities, are shown in Fig. 7 (b).

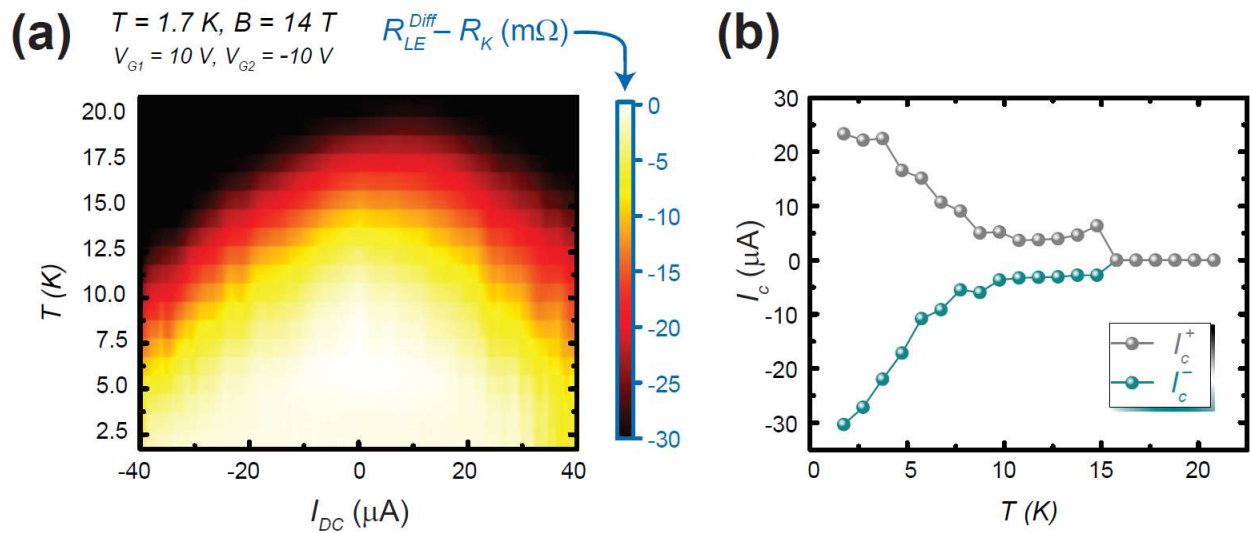

**Figure 7. Quantization as a function of temperature and critical current.** (a) The deviation of $R_{xx}$ from $R_K$, while the lower edge (LE) was quantized at $R_K$, was measured as a function of temperature and applied current. There is a clear boundary for the combination of current and temperature which permits full quantization at $R_K$. (**b**) The critical currents are plotted as a function of temperature for both polarities.

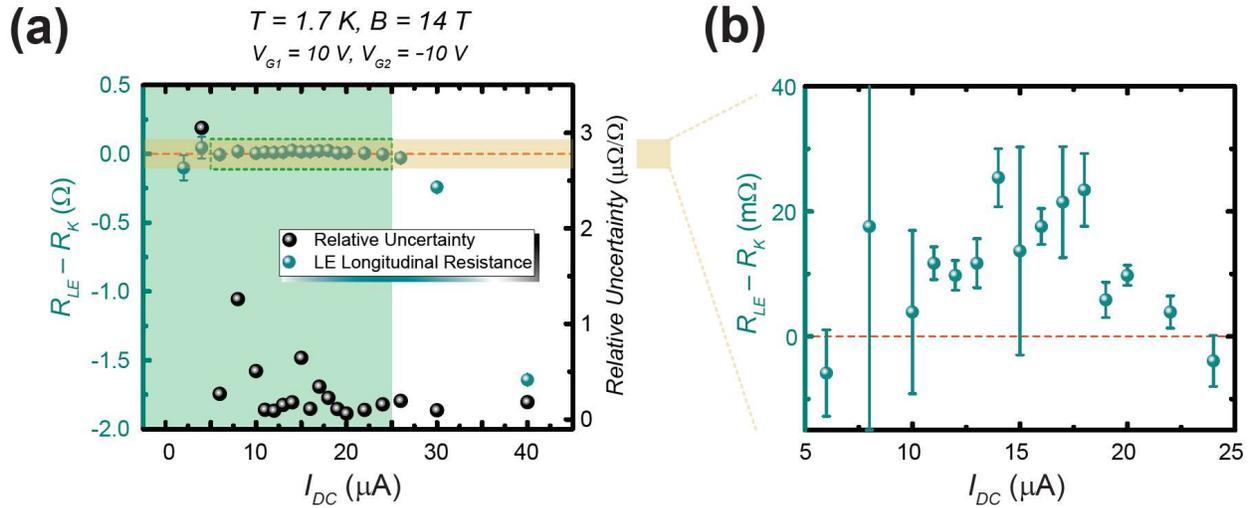

**Figure 8. Accuracy of quantization.** (**a**) The lower edge (LE) was compared against a 10 kΩ standard resistor with a DCC to give an accurate assessment of how well-quantized the device was when exhibiting a longitudinal resistance of $R_K$. The resistor was selected based on its traceability to a quantum resistance standard at the National Institute of Standards and Technology. The turquoise points show the DCC measurements as deviations from $R_K$ on the left axis and the relative uncertainties of those deviations with DC current. The relative uncertainties improve with increasing current, but the device loses its optimal quantization after the critical current of 24 μA. The shaded green area indicates the well-quantized region. (**b**) The beige shaded area in (**a**) is magnified to show the deviations' error bars as well as the reference to zero deviation, marked as an orange dashed line. The error bars represent a 1σ deviation from the mean, where each data point represents an average of a set of data taken at each value of current.

To assess the metrological usefulness of the *pn*J device, the quantity $R_{xx,\leq -R_K}$ was measured as a function of current used on the DCC ($I_{DC}$). The DCC, unlike its cryogenic counterparts, can provide turn-key resistance traceability for the most demanding applications, offering broader accessibility to these types of experiments.[35] This measurement places the device in a four-terminal bridge configuration against a 10 kΩ standard resistor, traceable to $R_K$, with the results shown in Fig. 8 (a). The measurement time for each data point with the DCC is 15 min, and the orange shaded region is magnified in Fig. 8 (b) to clarify the deviation of the DCC measurements with respect to zero. The data are displayed as turquoise points whose standard deviations (1s) are mostly smaller than the points. The right axis and its corresponding data, represented by black points, gives the relative uncertainty of each measurement as a function of $I_{DC}$. One important factor in resistance metrology is the level of precision one can achieve with such devices. In the case of this *p-n* junction, a precision of about $2 \times 10^7$ was achieved. Recall that the breakdown current is determined when $\frac{|\Delta R_{xx,LE}|}{R_K}$ is less than the relative uncertainty. The critical current for this case was 24 μA, and the region shaded in green encloses the current range one can potentially use for resistance metrology.

## Discussion

These data show that electrically programmable quantum resistance standards are feasible to build using *pn*Js as keystones. One can generalize these initial efforts for the future by proposing a specialized PQHR device, which will be described in more detail after explaining its most basic component, shown in Figure 9 (a). Consider *p*-type graphene formed into a device with $N$ regions with the following conditions: (1) Region 1 contains a single gate controlled by a single input voltage. (2) Region 2 contains two gates controlled by another single voltage input. (3) Region 3 contains four gates controlled by a third single voltage input. (4) If the device is extended to $N^{th}$ region, Region $N$ contains $2^{N-1}$ gates controlled by a single voltage input. We call a device with a total of $N$ regions an *N-bit* device, which can have a maximum of $2 * (2^N - 1)$ total *pn*Js. At each of the two ends of the graphene device, the usage of triple-series connection techniques is recommended to reduce the effect of contact and other resistances.[31]

For example, in the eventual proposed device, the largest device contains 8 regions (and is an *8-bit* device), where Region 8 contains 128 gates, all controlled by a single voltage input. In this case, there are 8 unique voltage sources for the 8 regions. All gates, when controlled by the voltage source, can be activated to shift the graphene carrier density to *n*-type, creating many *pn*Js. Let us denote when a region has its gates activated by defining the gate control parameter $b_i = 1 \lor 0$. If Region 3 has its gates activated, then $b_3 = 1$. We use this notation to count the total number of active *pn*Js in the given device: $b_N 2^{N-1} + b_{N-1} 2^{N-2} + b_{N-2} 2^{N-3} + \cdots + b_1 2^0$. Notice that when all $N$ regions are activated, the total number above is a geometric sum for the maximum we noted in the previous paragraph ($2 * (2^N - 1)$). Furthermore, we use the gate control parameter to form a binary string representing a device which has some (or all) of its $N$ regions activated: $b_N b_{N-1} b_{N-2} \cdots b_1$. In the *8-bit* device case, if all eight regions are activated, then the string is simply: 11111111. If only regions 1, 3, 5, and 7 are activated, then the string is: 01010101.

We now describe the second major component of this proposed device, shown in Figure 9 (b). An *N-bit* device can be used as one of many parallel devices, where $K$ indicates the number of parallel *N-bit* devices. The important thing to note here is that all $K$ devices share the gate control parameter, and thus the binary string is identical for all devices in parallel. If we have ten *8-bit* devices in parallel, then activating Region 3 would mean doing so for all ten devices, as the gates stretch over all devices. This is illustrated in Fig. 9 (b) as two *n*-type regions being identically activated over several devices by two long gates. This configuration allows an equal amount of current to flow in each of the $K$ branches and can be condensed by using two numbers to describe it: $N$, for the number of bits per isolated device, and $K$, the number of isolated devices in parallel. Let's now call this entire configuration an $(N, K)$ *module*.

The proposed specialized device for achieving seven decades of resistance is illustrated in Figure 9 (c). There are three stages of modules between the source and drain. Stage 1 contains four $(N, K)$ *modules* connected in parallel, stage 2 contains three $(N, K)$ *modules* in parallel, and stage 3 contains just one $(N, K)$ *module*. All three stages are connected in series with superconducting metal contacts.

When the numbers $N$ and $K$, along with the gate control parameter binary strings, are carefully selected, a wide range of resistance decades can be achieved. For example, Table 1 shows seven decades and their corresponding parameters. The example parameters are selected to ensure low deviation (< 1 μΩ/Ω) from decade values. The current flow is from source to drain and voltages can be measured using any two voltage probes shown in Fig. 9 (c), labelled *A, B, C* and *D*.

| Resistance | Stage 1 | Stage 2 | Stage 3 | Voltage probes used | Deviation from decade value |
|---|---|---|---|---|---|
| **100 Ω** | 00010010 00011001 00001001 00010001 | None | None | *A, B* | 0.714 μΩ/Ω |
| **1 kΩ** | 10001001 10001011 10000111 10010001 | None | None | *A, B* | 0.108 μΩ/Ω |
| **10 kΩ** | 00110001 00100111 00110011 00111100 | 011 010 010 | None | *A, C* | 14.8 nΩ/Ω |
| **100 kΩ** | 00111101 00111101 00111100 00111110 | 010 010 010 | 000000000011 | *A, D* | 0.043 nΩ/Ω |
| **1 MΩ** | 00101011 00100001 00110101 00110000 | 010 001 001 | 000000100110 | *A, D* | 0.0243 nΩ/Ω |
| **10 MΩ** | 01011110 01011001 01010011 01001100 | 110 101 110 | 000110000010 | *A, D* | 0.346 pΩ/Ω |
| **100 MΩ** | 01110111 10001001 01101101 01100111 | 100 011 011 | 111100100001 | *A, D* | 1.21 nΩ/Ω |

**Table 1. Possible resistance decades achievable with programmable standards.** The proposed device in Figure 9 (c) can be programmed with the values listed in this table to achieve very accurate values of seven decades of resistance. Each module in each stage is assigned a binary string. As long as the exact configuration is used, the measured voltage between the two probes will measure a near-exact decade value, to within a deviation defined in the rightmost column.

The proposed PQHR device in Fig. 9 (c) requires a total number of 53 voltage inputs to control 53 distinct regions (32 total regions in stage 1, 9 total regions in stage 2, and 12 total regions in stage 3). When fabricating such a device, the graphene widths will be identical within each stage, but will scale from stage to stage with the ratio of 1:7:3654, where graphene in first stage would be the narrowest. This would require, with a narrowest channel of 10 μm in stage 1 components, a stage 3 component width of about 3.7 cm.

**(a)**

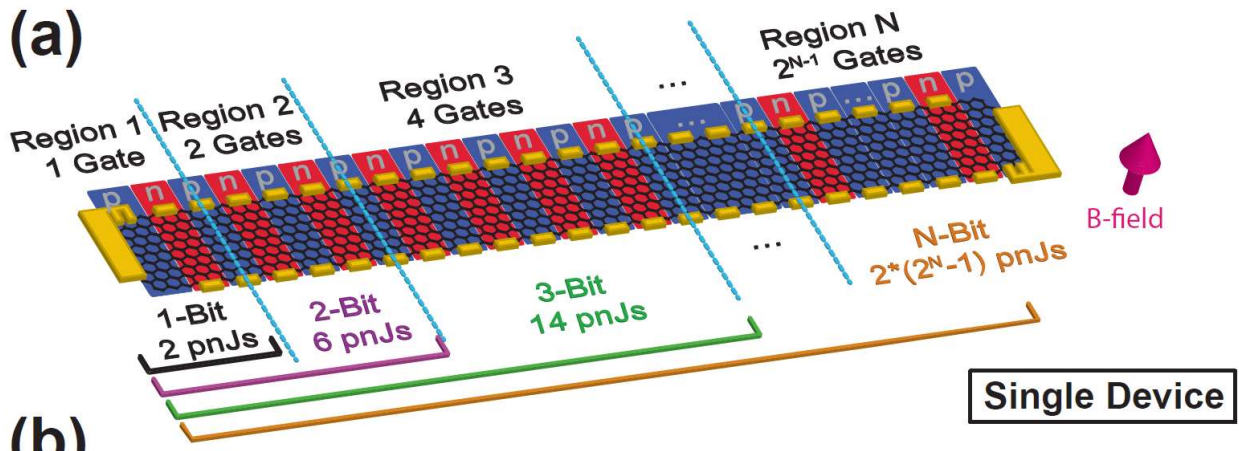

Region 1
1 Gate

Region 2
2 Gates

Region 3
4 Gates

...

Region N
$2^{N-1}$ Gates

B-field

1-Bit
2 pnJs

2-Bit
6 pnJs

3-Bit
14 pnJs

N-Bit
$2*(2^N-1)$ pnJs

**Single Device**

**(b)**

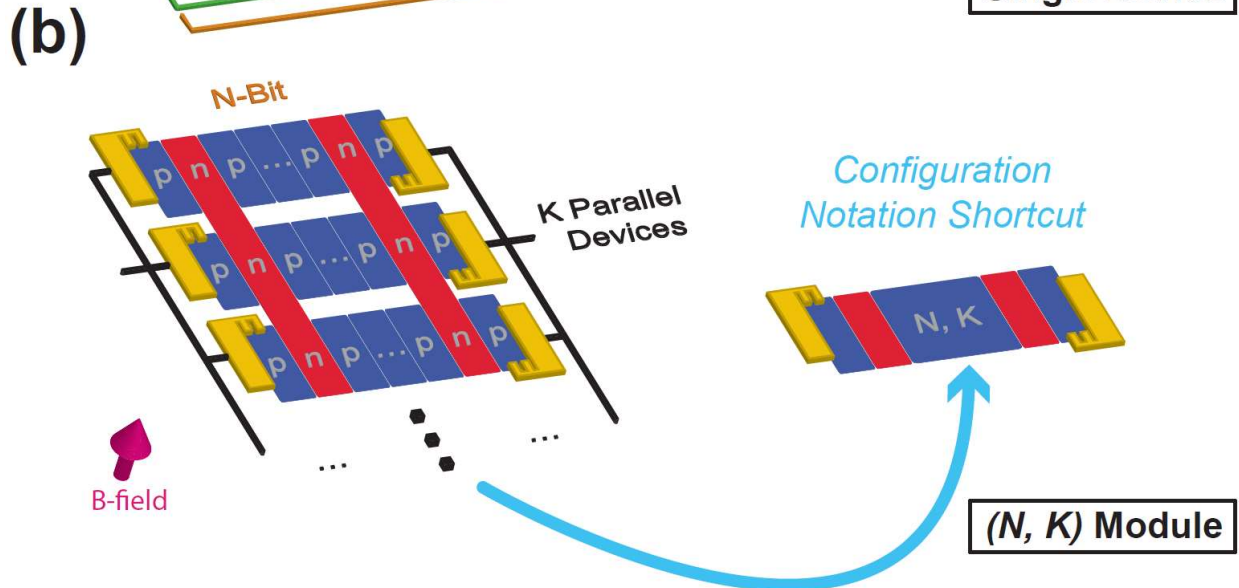

N-Bit

K Parallel
Devices

*Configuration
Notation Shortcut*

N, K

B-field

**(N, K) Module**

**(c)**

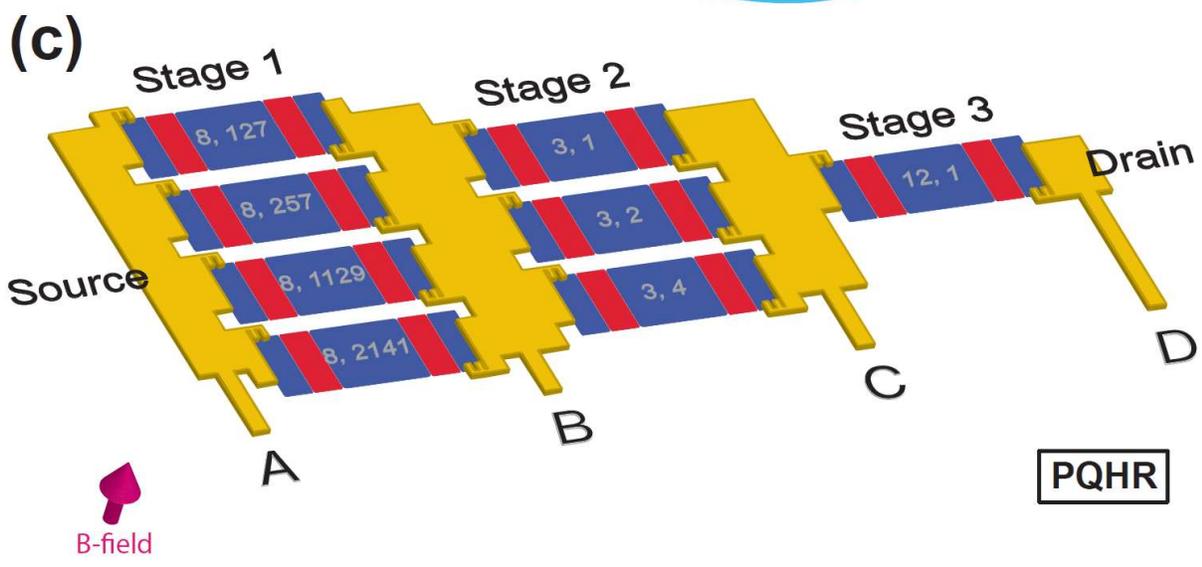

Stage 1

Stage 2

Stage 3

Drain

8, 127

8, 257

8, 1129

8, 2141

3, 1

3, 2

3, 4

12, 1

Source

D

A

B

C

B-field

**PQHR**

**Figure 9. Proposal of PQHR device for scalable and programmable standards.** (**a**) An *N-bit* device is illustrated showing how each region is defined and the maximum number of *pn*Js that can be used. (**b**) This device, when connected in parallel with $K$ copies of itself, becomes the foundation of the $(N, K)$ *module*. Each region has a set of gates that extend to all $K$ parallel branches. (**c**) The proposed device is illustrated and composed of eight $(N, K)$ *modules*, four of which are in parallel in stage 1, three of which are parallel in stage 2, and a lone *module* in stage 3. All three stages are connected in series and all connections and contacts are proposed to be superconducting metal to eliminate the contact resistance to the greatest possible extent. The *modules* in stage 1 are *8-bit* devices with more than 100 parallel copies per *module*, whereas the *modules* in stage 2 are *3-bit* devices with four or fewer parallel copies per *module*. Stage 3 is a single *12-bit* device with no additional parallel branches. These numbers for $(N, K)$ are required should one wish to reproduce the values in Table 1.

Comprehensive assessments of the magnetic field, carrier density, top gate voltage, critical current, temperature, and accurate quantization parameter spaces were conducted to determine the conditions required for an EG *pn*J device to be metrologically useful. In conclusion, after fabricating and measuring an EG *pn*J using *h*-BN as the gate dielectric, we demonstrated that accurate quantization of zero and $R_K$ can be exhibited by the device's longitudinal resistance with a relative uncertainty on the order of $10^{-7}$ as measured by a DCC resistance bridge. This work begins a new avenue in the production of quantum resistance standards and enables their accurate scaling for future metrology applications.

## Methods

**Sample growth.** The methods for growing high quality epitaxial graphene are well-reported.[5,6,11,36] EG is formed by sublimating Si atoms from the silicon face of SiC as part of an annealing process. Samples were grown on square SiC chips diced from on-axis $4H$-SiC(0001) semi-insulating wafers (CREE) [see additional information]. SiC chips were submerged in a 5:1 diluted solution of hydrofluoric acid and deionized water prior to the growth process. Chips were rinsed with deionized water and placed on a polished pyrolytic graphite substrate (SPI Glas 22) [see additional information]. Chips were processed with AZ5214E to utilize polymer-assisted sublimation growth techniques.[36] The silicon face was resting against the graphite in order for the gap between the two surfaces to create a diffusion barrier for escaping Si atoms. This configuration promotes homogeneous graphene growth conditions.[11] The annealing process was performed with a graphite-lined resistive-element furnace (Materials Research Furnaces Inc.) [see additional information], with heating and cooling rates of about 1.5 °C/s. The growth stage was performed in an ambient argon environment at 1900 °C.[5]

**Sample fabrication.** The grown EG was evaluated with confocal laser scanning and optical microscopy as an efficient way to identify large areas of successful growth.[37] Using photolithography, both the Hall bar geometry and the contact pads are fabricated in steps that are presented in detail in other works.[5,38] In summary, protective layers of Pd and Au are deposited on the EG to prevent organic contamination. While protected, the EG is etched into the desired device shape, with the final step being the removal of the protective layers from the Hall bar using a solution of 1:1 aqua regia to deionized water. To fabricate top gates, hexagonal boron nitride (*h*-BN) flakes are exfoliated onto polydimethylsiloxane (PDMS) and their quality and size are inspected with a dark field optical microscope and AFM. The PDMS slab carrying the selected *h*-BN flake is then mounted on a glass slide arm for positioning and alignment with the EG device, accurate to within a few μm, using a homemade transfer stage. The *h*-BN flake is slowly lowered towards the EG surface until contact is made. The stage, upon which the EG device rests, is then heated to 110 °C to dislodge the *h*-BN flake from the PDMS substrate. A standard electron beam lithography process is used to fabricate metal. All contacts had resistances of approximately 100 Ω in the QHE regime.

## Data Availability

All data generated or analysed during this study are included in this published article (and its Supplementary Information files).

## Acknowledgements

AFR would like to thank the National Research Council's Research Associateship Program for the opportunity. Work done by YY was supported by federal grant #70NANB12H185. The work of BYW at NIST was made possible by arrangement with Prof. CT Liang of National Taiwan University. Work done by JT was supported by federal grant #70NANB12H184. HYL would like to thank Theiss Research for the opportunity. The authors thank JA Stroscio for his assistance.

## Author contributions statement

JH performed sample fabrication, measurements, analyses, and managed the overall project direction. AFR assisted with sample characterization, data analyses, figure composition, and manuscript writing. MK, YY, and REE produced the EG growth. BYW, JT, and HYL assisted with sample fabrication and characterization. ARP, SUP, GRJ, MEK, and DGJ provided assistance with the bridge measurements and handling of standard resistors. KW and TT provided *h*-BN material for sample fabrication. REE and DBN contributed overall project ideas and consulting. All authors have approved the final manuscript.

# Additional information

**Supplementary information** accompanies this paper at http://www.nature.com/srep/

**Work prepared, in part, by employees of the United States Government:** Commercial equipment, instruments, and materials are identified in this paper in order to specify the experimental procedure adequately. Such identification is not intended to imply recommendation or endorsement by the National Institute of Standards and Technology or the United States government, nor is it intended to imply that the materials or equipment identified are necessarily the best available for the purpose.

**Use of the von Klitzing constant** $R_K$**:** All $R_K$ usage in the main text complies with the conventional electrical unit definition $R_{K-90} = 25812.807\Omega$. This value was agreed upon by the International Committee for Weights and Measures (CIPM) and adopted for international use in 1990.

**Competing interests:** All authors declare no competing interests.

# Supplementary information: Designs for programmable quantum resistance standards based on epitaxial graphene *p-n* junctions

Jiuning Hu[1,2*], Albert F. Rigosi[1*], Mattias Kruskopf[1,2], Yanfei Yang[1,2], Bi-Yi Wu[1,3], Jifa Tian[1,4], Alireza R. Panna[1], Hsin-Yen Lee[1,5], Shamith U. Payagala[1], George R. Jones[1], Marlin E. Kraft[1], Dean G. Jarrett[1], Kenji Watanabe[6], Takashi Taniguchi[6], Randolph E. Elmquist[1], and David B. Newell[1]

[1]Physical Measurement Laboratory, National Institute of Standards and Technology (NIST), Gaithersburg, MD 20899, USA
[2]Joint Quantum Institute, University of Maryland, College Park, MD 20742, USA
[3]Graduate Institute of Applied Physics, National Taiwan University, Taipei 10617, Taiwan
[4]Department of Physics and Astronomy, and Birck Nanotechnology Center, Purdue University, West Lafayette, Indiana 47907, USA
[5]Theiss Research, La Jolla, CA 92037, USA
[6]National Institute for Materials Science, 1-1 Namiki, Tsukuba 305-0044, Japan
*hujiuning@gmail.com

Contents

## 1. Different values of $R_{xx}$ as a function of two gate voltages and B-field

At the end of this document, we provide the dependence of the $R_{xx}$ maps as a function of two gate voltages. The maps are presented in successive rows of panels to show their dependence on the magnetic field and are arranged from left to right: UE ($R_{xx}$), LE ($R_{xx}$), UE ($R_{xy}$), LE ($R_{xy}$).

## 2. Dirac point determination for G1 and G2

To verify the expected symmetry seen between Figure 4 (a) and (c) in the main text, and to have a rigorous knowledge of the carrier density for any applied top gate voltages, the Dirac point was determined for both G1 and G2.

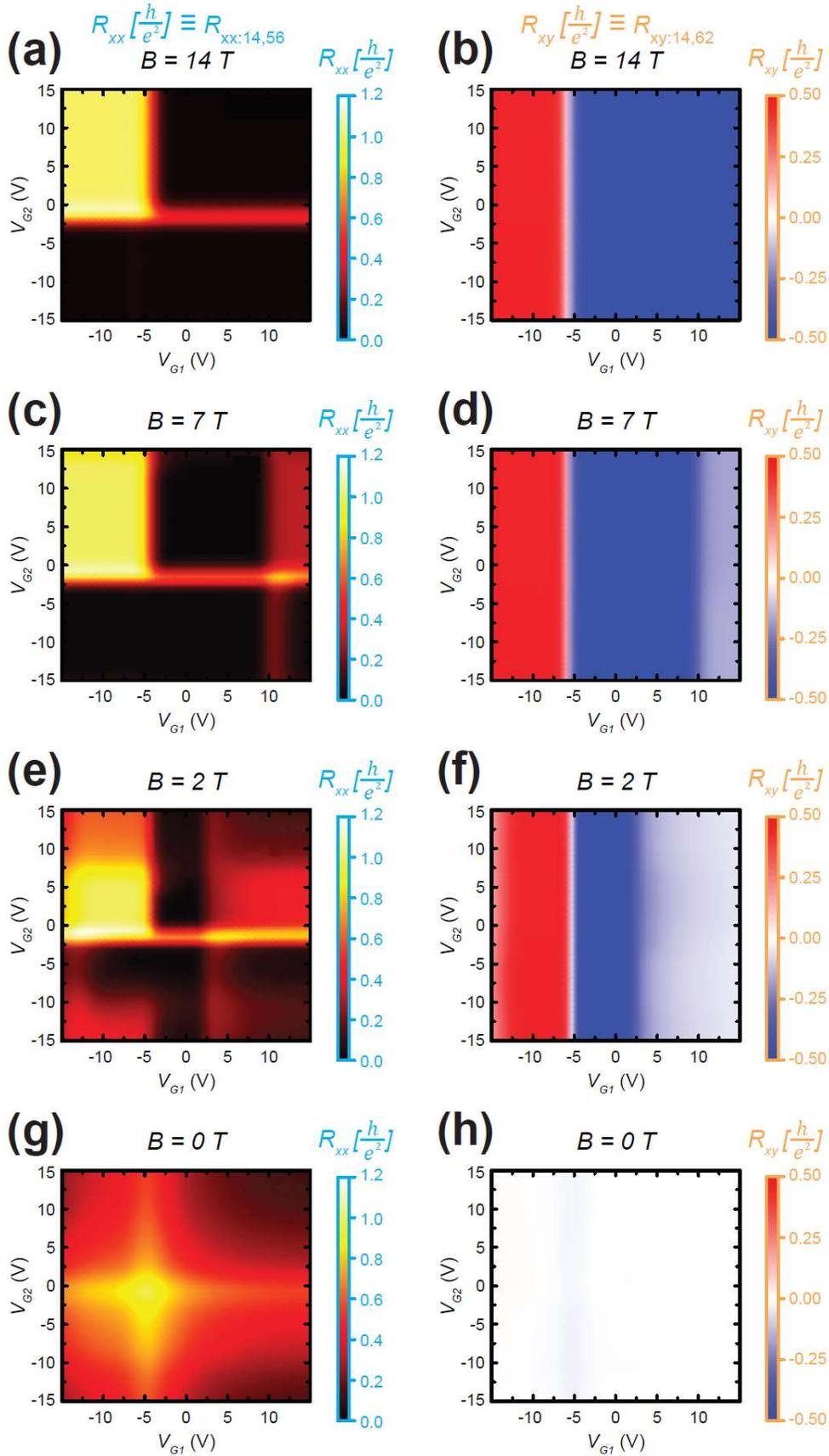

**Figure 1-SM.** The longitudinal and Hall resistances are displayed side-by-side with decreasing magnetic field towards the bottom of the figure. (a) and (b) are data collected at 14 T. The magnetic field in the subsequent panels are: (c) and (d) – 7 T, (e) and (f) 2 T, and (g) and (h) 0 T.

We include a few representative panels of how the B-field affects the resistance maps in Figure 1-SM. The Dirac points in Figure 1-SM are determined by horizontal or vertical cuts in the $R_{xx}$ maps, and for positive B fields, the results are averaged to obtain the final Dirac points. The off-diagonal cut across the Dirac points for the $R_{xx}$ maps at different B fields are plotted in Figure 3 in the main text.

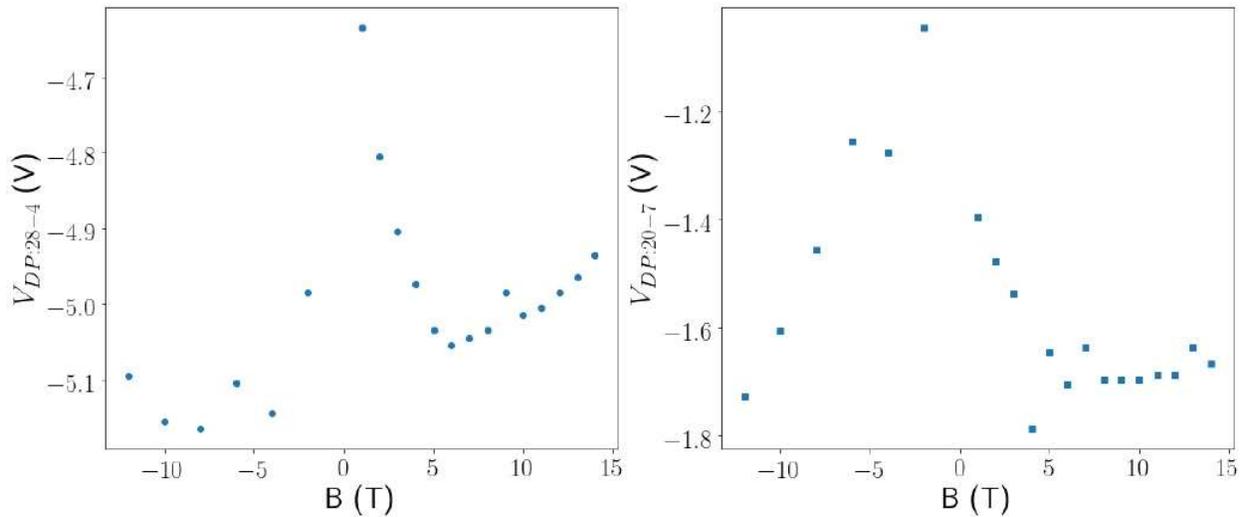

**Figure 2-SM.** (left) The data for the Dirac point of G1 are acquired at 1.7 K as a function of magnetic field, with the corresponding charge neutrality point being approximately at -1.68 V. (right) The Dirac point of G2 is determined with the same procedure, yielding the approximate charge neutrality point of -5.01 V.

After gaining the knowledge of the two Dirac points shown in Figure 2-SM, the expected points of symmetry were identified and marked in Figure 6 of the main text.